\documentclass[conference,10pt]{IEEEtran}
\usepackage{amssymb,amsmath,amsfonts,bm,epsfig,graphicx,theorem,latexsym}
\usepackage{rotating,setspace,latexsym,epsf,color,subfigure}
\usepackage{cite,float,epsf,authblk}

\newtheorem{Theorem}{Theorem}

\newenvironment{Proof}[1]{\medskip\par\noindent
{\bf Proof:\,}\,#1}{{\mbox{\,$\blacksquare$}\par}}
       
\def \n2{{N_0 \over 2}}

\def \h5{\hspace{0.5in}}

\begin{document}

\title{State Amplification and Masking While \\ Timely Updating \vspace{-0.1in}}

\author[1]{Omur Ozel}
\author[2]{Aylin Yener}
\author[3]{Sennur Ulukus}
\affil[1]{\normalsize Department of Electrical and Computer Engineering, George Washington University, Washington, DC}
\affil[2]{\normalsize Department of Electrical and Computer Engineering, The Ohio State University, Columbus, OH}
\affil[3]{\normalsize Department of Electrical and Computer Engineering, University of Maryland, College Park, MD}

\maketitle

\begin{abstract} 
In status update systems, multiple features carried by the status updating process require pursuit of objectives beyond timeliness measured by the age of information of updates. We consider such a problem where the transmitter sends status update messages through a noiseless binary energy harvesting channel that is equivalent to a timing channel. The transmitter aims to amplify or mask the energy state information that is carried in the updating process. The receiver extracts encoded information, infers the energy state sequence while maintaining timeliness of status updates. Consequently, the timings of the updates must be designed to control the message rate, the energy state uncertainty, and the age of information. We investigate this three-way trade-off between the achievable rate, the reduction in energy arrival state uncertainty, and the age of information, for zero and infinite battery cases. 
\end{abstract}

\section{Introduction}

Freshness of information is a crucial need for data transmitted in broad application domains such as internet of things and cyber physical systems, and is expected to be a part of next generation communications. For instance, optimal estimation of dynamical system states entails fresh information to be used where states represent physical variables in the system such as energy and channel strength. In addition to freshness, the timings of the updates typically carry additional information about the dynamics of the system over which the transmissions occur. Depending on the application, it may be preferable to amplify such information while maintaining data freshness. An instance of this problem arises in networks with state dependent channels where trends in the dynamics of the channel states are useful to take timely decisions about regime of operation. In the other extreme, it may be preferable to keep such information hidden from the receiver due to the privacy of sensitive data. These research problems are also connected to the emerging field of semantic communications, where the context of transmissions or transmitted messages themselves may embody or convey further information. 

In this paper, we investigate state amplification and state masking problems \cite{2014itw} in a timing channel derived from an energy harvesting channel under data freshness constraints. We focus on a binary energy harvesting channel where a status update is represented by a ``1'' symbol. The transmission of a ``1'' symbol improves freshness and at the same time reveals information about the presence of energy, a quantity that may be amplified or hidden. Equipped with a battery to buffer incoming energy, the timings of symbol transmissions determine information rates and data freshness simultaneously. Transmitter aims to amplify or mask the energy arrival state information that is carried in the updating process by designing the timings of updates subject to dynamic energy constraints and freshness of updates. We use the age of information (AoI) metric as a measure of data freshness; see recent surveys on AoI and its applications \cite{kosta2017age, SunSurvey, yates2020age}. 

Analysis and optimization of AoI in energy harvesting communications have been considered, e.g., in \cite{yates2015lazy, arafa-asilomar2017, wu2017optimal_ieee, baknina2018coded, farazi2018average,farazi2018age, feng2018optimal, bacinoglu2018achieving, 2018information, krikidis2019average, chen2019age, arafa2019ageinf, tunc2019optimal, arafa2019timely, bacinoglu2019optimal, zheng2019closed, rafiee2020active, ozel2020, sleem2020age, rafiee2021age, feng2021age, gindullina2021age, khorsandmanesh2021average, arafa2021timely} mainly focusing on average AoI minimization (with exceptions in \cite{ozel2020,rafiee2021age, khorsandmanesh2021average}) under offline and online knowledge of energy harvests. Additional directions that include energy harvesting constraints are cognitive radio \cite{sleng2019impact, leng2019minimizing, leng2019age, leng2019age2}, caching \cite{yates2017age, Bastopcu2020LineNetwork, kaswan-isit2021, pappas2020average}, remote estimation \cite{jaiswal2021minimization} and reinforcement learning \cite{ceran2019average,sleng2019age,abd2020reinforcement,leng2021actor,hatami2021aoi}. On another line of related research, information-theoretic limits in energy harvesting communications have been considered in \cite{ozel2012achieving, mao2017capacity, shaviv2016capacity, jog2016geometric, tutuncuoglu2017binary}.

\begin{figure}[t]
\centering{\includegraphics[width=0.86\columnwidth]{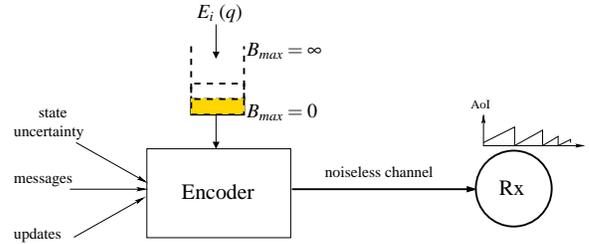}}
\caption{Model illustration for state amplification and masking in a noiseless binary energy harvesting channel with zero or infinite battery.}\vspace{-0.15in}
\label{fig:1} 
\end{figure}

When state amplification and masking problems are considered under AoI constraints, a three-way trade-off arises among age, rate and state uncertainty reduction. Making a transmission each time energy arrives is in line with maintaining data freshness and state amplification. On the other hand, randomized transmissions are more advantageous for message rates depending on the availability of battery. We examine this triple trade-off by using peak AoI and average AoI as age metrics, for zero and infinite battery cases. 

State amplification and state masking problems as well as age-rate trade-off have been considered separately in \cite{wu2017optimal_ieee, 2012information, 2014itw, 2018information}. In particular, for an infinite battery, \cite{wu2017optimal_ieee} shows that the encoder must apply a \emph{periodic} policy for optimal average AoI using the best-effort scheme in \cite{ozel2012achieving}. \cite{2018information} considers the age-rate trade-off in a binary energy harvesting channel but for an encoder with a unit sized battery inspired by the timing channel approach in \cite{tutuncuoglu2017binary,anantharam1996bits}. We observe in \cite{2018information} that sending information through a timing channel entails randomizing the timings of updates. On the other side, \cite{2012information,2014itw} consider state amplification and state masking problems in a binary energy harvesting channel for the zero and infinite sized battery cases. We observe in \cite{2012information,2014itw} that an infinite sized battery hides the energy arrival state information while sending information at the channel capacity limit. In contrast, there is a strict trade-off between message rate and state uncertainty reduction in the no battery case. In the present paper, we unify the notions in \cite{2012information, 2014itw, 2015information, 2018information} by defining and characterizing channel capacity under AoI constraints and then connecting it to the state amplification and masking problems. 

\section{System Model}
\label{sec:Model}

We consider the noiseless binary energy harvesting channel shown in Fig.~\ref{fig:1}. The transmitter sends status updates and an independent message simultaneously while also taking the history of energy arrival sequence into consideration to amplify or mask that information through its transmission decisions. The transmitter has either zero sized (i.e., no) or infinite sized battery, i.e., $B_{max}=0$ or $B_{max}=\infty$. In the case of zero battery the transmitter can transmit only at instants when energy arrives. In the case of infinite battery, the transmitter has the choice to transmit or save the energy in the battery for future use. In the latter case, the energy level in the battery at time $t$, denoted by $B_t$, evolves as 
\begin {align} 
B_{t+1} = B_t + E_t - X_t
\end{align} 
where $X_t \in \{0,1\}$ is the channel input. The transmission of ``1'' (resp. ``0'') symbol represents transmission (resp. absence) of an update and it costs one (resp. zero) unit of energy. Energy arrivals are distributed according to an i.i.d.~Bernoulli distribution with $\mathbb{P}[E_t=1]=1-\mathbb{P}[E_t=0]=q$.  

\begin{figure}[t]
\centering{\includegraphics[width=0.9\columnwidth]{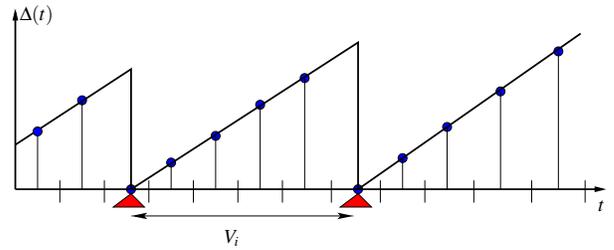}}
\caption{Evolution of AoI in discrete time.}
\label{fig:2} 
\vspace*{-0.25cm}
\end{figure}

The instantaneous AoI is given by $A(t)=t-u(t)$ where $u(t)$ is the time stamp of the latest received ``1'' symbol and $t$ is the current time. An example evolution of AoI in discrete time $t$ is shown in Fig.~\ref{fig:2}, where the red triangles are the slots at which updates are transmitted. The average peak AoI is
\begin{align}
\label{peakaoi}
A_p =\limsup_{n\rightarrow \infty}  \frac{1}{n}\mathbb{E}\left[\sum_{i=1}^{n}V_i \right]-1
\end{align}
where $V_i$ is the duration between two consecutive updates as shown in Fig.~\ref{fig:2}. Similarly, the average AoI is
\begin{align}
\label{averageaoi}
A_a =\limsup_{n\rightarrow \infty}  \mathbb{E}\left[\frac{\sum_{i=1}^{n}V_i^2}{2\sum_{j=1}^{n}V_i}\right]-\frac{1}{2}
\end{align}
The subtractions of 1 and $\frac{1}{2}$ in (\ref{peakaoi}) and (\ref{averageaoi}) are needed to offset the value of $A(t)$ with respect to $V_i$. Nevertheless, they are inconsequential for our analysis and we omit them in the rest of the paper. These age expressions reduce to the following single letter forms under stationary transmission policies,
\begin{align} 
A_p = \mathbb{E}\left[V\right], \qquad A_a = \frac{\mathbb{E}\left[V^2\right]}{2\mathbb{E}\left[V\right]}.
\end{align}

We are interested in the information the decoder can learn about the energy arrival process $E^n$. There are $2^{H(E^n)}$ possible energy arrival sequences, since $2^{H(E^n)}$ is the size of the typical set for $E^n$ where $H(\cdot)$ denotes the entropy. When the symbol sequence $Y^n$ is received, which is identical to the transmitted sequence $X^n$ due to the noiseless channel, the decoder can reduce the size of this list to $2^{H(E^n|Y^n)}$. Hence, the reduction in the energy state uncertainty (ESU) for the decoder is 
\begin{align}
\Delta = \frac{1}{n} \left(H(E^n) - H(E^n|Y^n)\right) = \frac{1}{n} I(E^n;Y^n).
\end{align}

If the objective is to amplify the information in $E^n$, we aim to maximize $\Delta$ (i.e., we find an upper bound on it). If the objective is to mask that information, we minimize $\Delta$ (i.e., we find a lower bound on it). These optimizations are performed subject to a maximum message rate constraint. We will use $\Delta_a$ and $\Delta_m$ to denote the association with \emph{amplification} and \emph{masking} objectives. We briefly explain state amplification and state masking problems next. 

\emph{State amplification:} In this problem, the encoder wishes the decoder to obtain as much information as possible about the energy harvesting process $E^n$, i.e., maximize $\Delta_a$, while reliably conveying a message with some rate $R$. This problem is first considered in \cite{kim2008state}, where the achievable message rates and state amplification rates are shown to satisfy
\begin{align}
\label{eqn_model_amp_region_a}
	R \leq I(U;Y), \quad \Delta_a \leq H(S), \quad	R+ \Delta_a \leq I(X,S;Y),
\end{align}
for a memoryless channel with state $S$ known causally at the transmitter. Here, $U$ is an auxiliary random variable yielding the joint distribution $p(s)p(u)p(x|u,s)p(y|x,s)$. 

\emph{State masking:} In this problem, the objective is to determine the minimum information $\Delta_m$ that must be revealed to the decoder about the state in order to achieve some rate $R$. The achievable $(R,\Delta_m)$ are obtained by the union of the regions 
\begin{align}
\label{eqn_model_mask_region_a}
R \leq I(U;Y), \quad \Delta_m \geq I(S;Y|U),
\end{align}
for causally available state at the encoder in a memoryless channel \cite{merhav2007information}. Again, $U$ is an auxiliary random variable. 

We denote the AoI-rate-ESU trade-off by the triple $(A(R,\Delta),R,\Delta)$, where $R$ is the achievable rate, $\Delta$ is the ESU reduction and $A(R,\Delta)$ is the minimum achievable average peak AoI and/or average AoI given that a message rate of $R$ and a energy state uncertainty reduction $\Delta$ is achievable. To analyze this triple trade-off, we will consider constraints on the average AoI and average peak AoI.

\section{Infinite Battery Case: $B_{max}=\infty$} \label{sect_emaxinf}

\subsection{Minimum AoI and Channel Capacity}
Let us first revisit the problem of minimizing AoI in the infinite battery case, i.e., $B_{max}=\infty$. Recall that \cite{wu2017optimal_ieee} considers a version of this problem where the time is continuous and the problem is posed for average AoI only. It is shown in \cite{wu2017optimal_ieee} that the transmitter must apply a \emph{periodic} policy for optimal average AoI using the best-effort scheme in \cite{ozel2012achieving}. In our problem setting, time is discrete. In order to accommodate the discreteness in time, we consider probabilistic periods of $\lfloor\frac{1}{q}\rfloor$ and $\lceil\frac{1}{q}\rceil$ so that $\mathbb{E}[V]=\frac{1}{q}$. If $\frac{1}{q}$ is an integer, the scheme in \cite{wu2017optimal_ieee} with period $\frac{1}{q}$ is optimal and the minimum average AoI is $A_a^{min,\infty}=\frac{1}{2q}$. If $\frac{1}{q}$ is not an integer, transmitting with probability $g_f=\lfloor\frac{1}{q}\rfloor+1-\frac{1}{q}$ at period $\lfloor\frac{1}{q}\rfloor$ and with probability $1-g_f$ at period $\lceil\frac{1}{q}\rceil$ achieves the optimal average AoI, which is $A_a^{min,\infty}=\frac{1}{2q}+\frac{qg_f(1-g_f)}{2}$. We observe that the discreteness of time causes an additive cost on the optimal average AoI.

We also note that the minimum average peak AoI in this case is $A_p^{min,\infty}=\frac{1}{q}$. We have $\mathbb{E}[V]\geq \frac{1}{q}$ due to the long term energy causality which imposes an average energy expenditure that is less than the average energy arrival rate. We can achieve as small an $\mathbb{E}[V]$ as $\frac{1}{q}$ by a best-effort or save-and-transmit type scheme \cite{ozel2012achieving}. Evidently, we can describe infinitely many schemes that achieve minimum peak AoI. It is interesting to observe that the scheme that minimizes average AoI is included among those that minimize peak AoI. Still, this result represents just a singular point. When state amplification and masking problems are considered, the behavior of the problem may differ under average and peak AoI constraints.

Toward that end, we next consider channel capacity under peak AoI and average AoI constraints:
\begin{align}\label{agec}
A_p \leq c_p, \quad A_a \leq c_a,
\end{align}
for some $A_a^{min,\infty} \leq c_a $ and $\frac{1}{q} \leq c_p$. For $c_a<A_a^{min,\infty}$ or $c_p < \frac{1}{q}$, the problem is infeasible.

\begin{Theorem}
The channel capacity under the two age constraints in (\ref{agec}), denoted by $C(c_p,c_a,q)$, is obtained in single-letter form by solving the following optimization problem: 
\begin{align}
\label{eq_best_ach0}
\max_{p(v)} \quad & \frac{H(V)}{\mathbb{E}[V]} \nonumber\\
\mbox{s.t.} \quad & \mathbb{E}[V^2] \leq 2c_a\mathbb{E}[V] \nonumber \\
& \frac{1}{q} \leq \mathbb{E}[V] \leq c_p.
\end{align}
\end{Theorem}

\begin{Proof}
We only give a sketch of the proof; the details are deferred to a longer version of this work. We first formalize the notion of timing channel and its relation to the usual channel. We then invoke \cite[Lemma 1]{tutuncuoglu2017binary} to prove equality of capacities in the timing channel and the usual channel. In particular, since the channel is noiseless, encoding and decoding information using the number of channel uses between two consecutive 1 symbol transmissions is equivalent to performing these operations in the usual channel. We finally use the save-and-transmit scheme in \cite{ozel2012achieving} with an average interval length larger than or equal to $\frac{1}{q}+\epsilon$ where $\epsilon$ is negligibly small. Recall that with an infinite battery, \cite{ozel2012achieving} shows that the save-and-transmit scheme achieves the capacity with the corresponding average transmit power. 
\end{Proof}

Note that the capacity under only a peak AoI constraint is:
\begin{align}
C(c_p,\infty,q) = H_b(\alpha)
\end{align}
where $\alpha=\min \left(q, \tfrac{1}{2} \right)$ if $c_p>2$ and otherwise $\alpha=\tfrac{1}{c_p}$ if $c_p<2$ is feasible and $H_b(\cdot)$ is the binary entropy function. That is, the capacity is achieved by geometric distributed $V$ under only peak AoI constraint. We observe that a strict trade-off between peak AoI and channel capacity arises if $q>\frac{1}{2}$ despite the fact that they are both achieved by geometric distributed $V$. In particular, achieving the channel capacity requires a codebook generation with equally distributed transmission of 0s and 1s that leaves a surplus of 1s in the battery. However, achieving optimal peak AoI requires transmission as soon as a new energy arrives. We also note that a closed form expression for $C(\infty,c_a,q)$ is not available since the second order moment constraint imposed by the average AoI is harder to analyze.

\subsection{State Amplification and Masking under AoI Constraints}
We are now ready to consider state amplification and state masking problems. The following theorem presents state amplification region for this channel under AoI constraints.

\begin{Theorem} \label{lem_emaxinf_amp}
The exact $(R,\Delta_a)$ region for the binary energy harvesting channel with an infinite sized battery at the transmitter under peak AoI constraint $c_p$ and average AoI constraint $c_a$ is:
\begin{align}
	\label{eqn_emaxinf_lemma_amp_1}
	R+\Delta_a \leq C(c_p,c_a,q), \quad 0 \leq \Delta_a \leq H_b(q).
\end{align}
\end{Theorem}

\begin{Proof}
	We only give a sketch of the proof. The achievability of these $(R,\Delta_a)$ pairs follows by the save-and-transmit scheme and by compressing the $E^n$ sequence using a block Markov encoding and sending it as a part of the message to swap any portion of the message rate $R$ with $\Delta_a$, provided that this portion does not exceed $H(q)$. The converse follows from similar lines to the converse argument in \cite[Lemma 1]{2014itw} along with an equivalence argument in \cite[Lemma 1]{tutuncuoglu2017binary} for the timing channel and the usual channel, unifying the message rate and the state uncertainty reduction.
\end{Proof}

In light of Theorem 1, it suffices to find $C(c_p,c_a,q)$ by solving (\ref{eq_best_ach0}) to determine feasible $(R,\Delta_a)$ pairs. We observe that the inequality $0 \leq \Delta_a \leq H(q)$ is redundant if $q \leq \frac{1}{2}$. In this case, $R$ and $\Delta_a$ can be swapped up to the age constrained capacity limit.

For the state masking problem, as shown in \cite{2012information}, since $(R,\Delta_m)=(C(c_p,c_a,q),0)$ is achievable, perfect masking of the state $E^n$ is possible using the save-and-transmit scheme for any achievable rate $R$. Hence, we have $\Delta_m \geq 0$ as the state masking lower bound for any $c_a$, $c_p$ and $q$.

\subsection{Computing the Solution of (\ref{eq_best_ach0})}

We now focus solving (\ref{eq_best_ach0}). We write (\ref{eq_best_ach0}) as,
\begin{align}
\label{eq_best_ach}
\max_{p(v),\frac{1}{q} \leq K \leq a_p} \quad & \frac{H(V)}{K} \nonumber\\
\mbox{s.t.} \quad & \mathbb{E}[V^2] \leq 2a_a\mathbb{E}[V] \nonumber \\
& \mathbb{E}[V] = K.
\end{align}

For a fixed $K$, the problem in (\ref{eq_best_ach}) is a convex problem in the PMF $p(v)$ and can be solved approximately in CVX using a finite dimensional $p(v)$. Then, to obtain the trade-off region, we sweep over all possible values of $K$ (i.e., all possible values of the peak AoI). We are guaranteed to observe a monotone decrease in the rate after a certain level of $K$ and find a solution due to concavity. The solution for (\ref{eq_best_ach}) may not have a closed form; hence, we propose the following simpler policies with closed form expressions.

\subsection{Wait-and-Transmit and Zero-Wait Policies}
In the wait-and-transmit policy, the transmitter waits until a threshold $\omega \geq 1$ slots since the last update ($\omega=1$ means no waiting). Then, it transmits with probability $p$. That is, the transmitter chooses $p(v)$ as follows
\begin{align}
p(v)=
\begin{cases}
0, & v<\omega\\
g (1-g)^{v-\omega}, &  v\geq \omega,
\end{cases}
\end{align}
for $v=1,2,\cdots$.
In this case, $g$ and $\omega$ are the variables over which the optimization is performed. The achieved information rate as a function of $g$ and $\omega$ is 
\begin{align}
R=\frac{\frac{H_b(g)}{g}}{c - 1 + \frac{1}{g}}.
\end{align}
Similarly, we calculate the average AoI with this policy as $
A_a = \frac{\mathbb{E}[V^2] }{2\mathbb{E}[V]}$, where 
\begin{align}
\mathbb{E}[V^2]= (\omega-1)^2 + 2(\omega-1) \frac{1}{g} + \frac{2-g}{g^2}.
\end{align}
Further, the peak AoI is equal to $A_p=\mathbb{E}[V]=\omega-1+\frac{1}{g}$. We then search for the optimal $g$ and $\omega$ subject to constraints on average AoI $c_a$ and peak AoI $c_p$. In particular note that $\omega - 1 + \frac{1}{g} \geq \frac{1}{q}$ due to the average energy causality.

We observe that waiting is useful only for average AoI minimization. Otherwise, increasing $\omega$ decreases $R$ and increases $A_p$. We expect therefore that waiting will help when $c_a$ is close to $A_a^{min,\infty}$. We will investigate the regimes in which waiting is advantageous numerically. 

Next, we consider the zero-wait transmission policy. This policy is a version of the wait-and-transmit policy where there is no waiting after an update transmission, i.e., $\omega=1$. This policy transmits with probability $g$ right away. Hence, the best achievable rate by this policy is obtained by solving
\begin{align}
\max_{\frac{1}{c_p} \leq g \leq q} \quad & H_b(g)     \nonumber \\
\mbox{s.t.} \quad & \mathbb{E}[V^2] \leq 2 c_a \mathbb{E}[V], 
\end{align}
given $c_a$ and $c_p$ where $\mathbb{E}[V^2]=\frac{2-g}{g^2}$ and $\mathbb{E}[V]=\frac{1}{g}$. Hence, the constraint above translates into a constraint as $g \geq \frac{2}{1+2c_a}$. Note that the condition $g \leq q$ guarantees the average energy causality. This optimization problem involves only a single variable $g$ and the solution is obtained by a search.

We note that the state masking performances of the proposed transmission policies remain perfect (i.e., $\Delta_m \geq 0$ is achievable) since they can be implemented by using a save-and-transmit approach despite their rate performances that fall below the channel capacity.

\section{No Battery Case: $B_{max}=0$} \label{sect_emax0}

For $B_{max}=0$, the energy available at channel use $t$, $E_t$, is the state of the channel, which is i.i.d.~and memoryless. Therefore, all objectives including message rate, state uncertainty reduction and AoI are optimized by a stationary and memoryless policy with respect to $E_t$. In particular, the results of \cite{kim2008state} are applicable in this case. Given the state, $E_t\in\{0,1\}$, the input $X_t\in\{0,1\}$, and the restriction $X_t \leq E_t$, there are two feasible mappings from $E$ to $X$. We denote them as $U=(X,\bar{X})$, where $X$ is the channel input when $E=0$ and $\bar{X}$ is the channel input when $E=1$. The two feasible strategies are $(0,0)$ and $(0,1)$. For an encoding strategy with $\mathbb{P}[U=(0,1)]=p$, the exact $(R,\Delta_a)$ region for state amplification is obtained from (\ref{eqn_model_amp_region_a}) as
\begin{align}
\label{eqn_emax0_amp_a}
\hspace{-0.1in} R \leq H_b(pq)-pH_b(q), \Delta_a  \leq H_b(q), R+ \Delta_a  \leq H_b(pq)
\end{align}

Next, we use \cite{merhav2007information} to characterize the exact $(R,\Delta_m)$ region for state masking with zero battery as 
\begin{align}
\label{eqn_emax0_mask}
R \leq H_b(pq)-pH_b(q), \quad \Delta_m &\geq pH_b(q)
\end{align}

The average peak AoI and average AoI in this case are:
\begin{align}
A_p = \frac{1}{pq}, \quad A_a = \frac{2-pq}{2pq}
\end{align} 
We note that $A_p$ and $A_a$ are minimized by transmitting each time an energy arrival occurs, i.e., by setting $p=1$. Then, we have the corresponding minimum values $A_p^{min,0} = \frac{1}{q}$ and $A_a^{min,0} = \frac{2-q}{2q}$, where $0$ in the superscript denotes the correspondence to zero battery. In contrast, note that sending messages along with state amplification or masking entails randomized transmissions that depart from age minimal transmission. We will obtain the triple trade-off among AoI, rate and state uncertainty reduction in our numerical results by sweeping over possible transmission probabilities.

\section{Numerical Results} \label{sect_numerical}

\begin{figure}[t]
\centering{\includegraphics[width=0.97\linewidth]{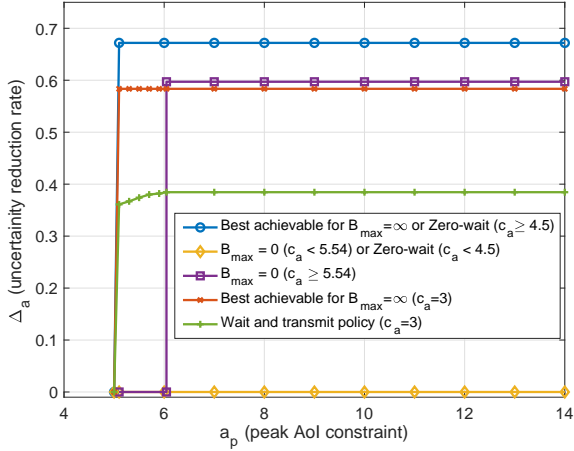}}
\caption{The state amplification performance $\Delta_a$ with respect to the peak AoI constraint $c_p$ for $q=0.2$ and $R\geq 0.05$ and various $c_a$.} 
\label{fig_results_amp}
\vspace*{-0.1cm}
\end{figure}

\begin{figure}[t]
\centering{\includegraphics[width=0.97\linewidth]{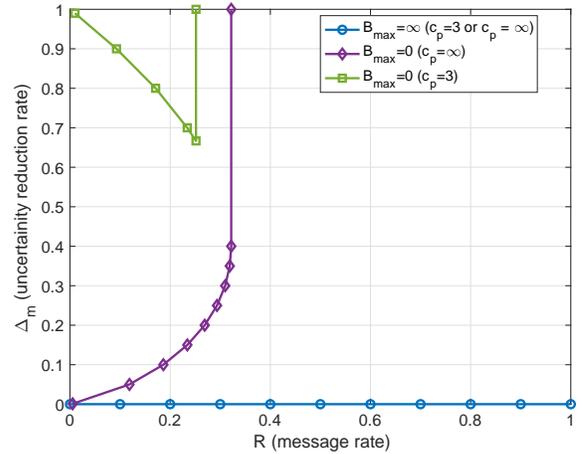}}
\caption{The state masking performance $\Delta_m$ with respect to $R$ for $q=0.5$ with various $c_p$.} 
\label{fig_results_mask}
\vspace*{-0.1cm}
\end{figure}

In this section, we provide numerical results that illustrate our theoretical results for several operating points of interest. The state amplification results are plotted in Fig.~\ref{fig_results_amp}, and the state masking results are plotted in Fig.~\ref{fig_results_mask}. 

In Fig.~\ref{fig_results_amp}, we observe the best achievable $\Delta_a$ with respect to the peak AoI constraint $c_p$ for various average AoI constraints $c_a$. We use $q=0.2$ for a low energy regime for which we have $A_p^{min,0}=A_p^{min,\infty}=5$, $A_a^{min,0}=4.5$ and $A_a^{min,\infty}=2.5$. We set the message rate $R \geq 0.05$ as the minimum required. We observe that when average AoI constraint $c_a$ is loose, the performance of zero-wait policy achieves identical level of $\Delta_a$ to the capacity $C(c_p,c_a,q)$ minus 0.05. This actually holds true for any $q\leq \frac{1}{2}$ and $c_p \geq 4.5$; and best wait-and-transmit policy reduces to zero-wait. Note that this best achievable $\Delta_a$ level is significantly above the performance that can be achieved by $B_{max}=0$. In particular, when $c_p$ and/or $c_a$ are too restrictive to allow a rate $R \geq 0.05$ under $B_{max}=0$, we have $\Delta_a=0$. This is true especially when $c_a < 5.54$ and/or $c_p < 6.04$, which enforces transmission probability to be $p > 0.82$ under $B_{max}=0$ but then the rate $R$ falls below the required level. On the other hand, $c_a \geq 5.54$ and $c_p \geq 6.04$ are sufficient to guarantee a transmission probability $p=0.82$ that enables $R=0.05$, minimal AoI and maximum $\Delta_a$ of 0.59. Similarly, for zero-wait scheme $c_a < 4.5$ is infeasible and hence $\Delta_a=0$. When $2.5 < c_a < 4.5$, it is possible to obtain a non-zero rate $R$ with $B_{max}=\infty$ but this rate is below the required level when $2.5 < c_a < 2.52$. Finally, $c_a < 2.5$ is infeasible and $c_a=2.5$ is possible only by a periodic transmission scheme, which achieves zero rate due to the deterministic transmission intervals. In all cases, the unlimited battery allows significant improvement on top of the zero battery by accumulating energy and exchanging a portion of the achievable message rate for state amplification. In particular, we observe that even under stringent AoI constraints, wait-and-transmit policy can achieve a good improvement in $\Delta_a$ with respect to $B_{max}=0$ and the best achievable scheme provides even further improvement with respect to the threshold based scheme. We also note that the plots in Fig.~\ref{fig_results_amp} are indicative of a zero-one law with respect to the peak AoI constraint, which we aim to analytically explore in future work.

For the state masking problem, we compare the effects of stringent AoI and relaxed AoI constraints on the achievable $(R,\Delta_m)$ pairs. We have already observed that it is always achievable to have $\Delta_m \geq 0$ with $B_{max}=\infty$ despite potential differences in the maximum achievable rate by different transmission policies for any $c_p$, $c_a$ and $q$. It then remains to observe the effect of AoI constraints on $\Delta_m$ with $B_{max}=0$. We set $q=0.5$ for which we have $A_p^{min,0}=2$. We will examine $c_p=3$ and $c_p=\infty$ in Fig.~\ref{fig_results_mask}. These constraints translate to $p \geq 0.66$ and $p \geq 0$, respectively. We do not put additional average AoI constraint $c_a$ (i.e., we set $c_a=\infty$) due to the fact that its effect on the transmission policy is identical to that of $c_p$ for the case of $B_{max}=0$. Indeed, in this case, if average AoI constraint is binding, we can find an equivalent $c_p$ that makes that constraint redundant. This is despite the effect of $c_a$ on the capacity for $B_{max}=\infty$. When $c_a=\infty$, we have $C(c_p,c_a,q)=1$. We observe that the $(R,\Delta_m)$ region is the small triangular one when $c_p=3$ and it grows significantly when $c_p=\infty$ and ultimately becomes the whole region when $B_{max}=\infty$. For $B_{max}=0$, the behavior of the lower bound on $\Delta_m$ turns opposite when the AoI constraint transitions from stringent to relaxed. If it is stringent, we can mask the energy arrival state easier while transmitting with higher message rate $R$. On the other hand, when AoI constraint is relaxed, transmitting with a higher rate entails more information leakage about the energy arrival state. We also note that this general trend is observed irrespective of the energy arrival rate $q$.    

\section{Conclusion} \label{sec:conc}

We considered a new status updating problem where the update times are controlled to achieve state amplification or state masking in a binary energy harvesting channel under data freshness constraints. We focused on zero battery and infinite battery cases and determined explicit capacity results that govern the triple trade-off among age, message rate and energy arrival state uncertainty reduction, in zero and infinite battery cases. We proposed additional schemes of practical interest in the infinite battery regime, and provided numerical results comparing state amplification and masking performances. Future work includes investigating this trade-off for arbitrary finite battery sizes.

\end{document}